\documentclass[%
 reprint,
superscriptaddress,
 amsmath,amssymb,
 aps,
]{revtex4-1}

\usepackage{graphicx}
\usepackage{dcolumn}
\usepackage{bm}
\usepackage{color}

\newcommand{\+}{\uparrow}
\renewcommand{\-}{\downarrow}
\renewcommand{\>}{\rangle}
\newcommand{\<}{\langle}

\newcommand{\chatdag}{\hat{c}^{\dag}}
\newcommand{\chat}{\hat{c}}

\newcommand{\fhatdag}{\hat{f}^{\dag}}
\newcommand{\fhat}{\hat{f}}

\newcommand{\zhat}{\hat{z}}

\newcommand{\Dd}{\mathcal{D}}
\newcommand{\Gg}{\mathcal{G}}
\newcommand{\Hh}{\mathcal{H}}

\newcommand{\Mm}{\mathcal{M}}

\newcommand{\Zz}{\mathcal{Z}}

\newcommand{\Ss}{\mathcal{S}}
\newcommand{\Tt}{\mathcal{T}}

\begin{document}

\title{Many-particle Majorana bound states: derivation and signatures in superconducting double quantum dots} 

\author{Thomas E. O'Brien}
\affiliation{Perimeter Institute for Theoretical Physics, Waterloo, N2L 2Y5 Ontario, Canada}
\affiliation{School of Mathematics and Physics, University of Queensland, Brisbane, 4072 Queensland, Australia}%
 \author{Anthony R. Wright}%
 \affiliation{School of Mathematics and Physics, University of Queensland, Brisbane, 4072 Queensland, Australia}%
\author{Menno Veldhorst}
\email{m.veldhorst@unsw.edu.au}
\affiliation{
ARC Centre of Excellence for Quantum Computation and Communication Technology, School of Electrical Engineering \& Telecommunications, The University of New South Wales, Sydney 2052, Australia}%

\date{\today}

\begin{abstract}
We consider two interacting quantum dots coupled by standard \textit{s-wave} superconductors. We derive an effective Hamiltonian, and show that over a wide parameter range a degenerate ground state can be obtained. An exotic form of Majorana bound states are supported at these degeneracies, and the system can be adiabatically tuned to a limit in which it is equivalent to the one-dimensional wire model of Kitaev. We give the form of a Majorana bound state in this system in the strong interaction limit in the many-particle picture. We also study the Josephson current in this system, and demonstrate that a double slit-like pattern emerges in the presence of an extra magnetic field. This pattern is shown to disappear with increasing interaction strength, which is due to the current being carried by chargeless Majorana bound states.
\end{abstract}

\pacs{Valid PACS appear here}
\maketitle

\section{\label{sec:Introduction}Introduction}
The idea of protected ground state degeneracies in condensed matter systems has attracted much interest recently. It is proposed that in a $p$-wave superconductor where both spin and particle-hole degeneracy are absent, a zero-energy state called the Majorana bound state can appear. This elusive quasiparticle obeys non-Abelian statistics \cite{Ivanov}, and thus could serve as a qubit building block for topological quantum computation \cite{Freedman,SimonReview,AliceaReview}. While $p$-wave superconductors are rare, nanotechnology opens the possibility to design this unconventional superconducting state using suitable combinations of materials.

With these prospects in mind, the hunt for Majorana bound states is stronger than ever, and there is much experimental activity to realize proposed schemes that contain these particles. Recent work includes superconductor-topological insulator systems \cite{FuKane2009} and semiconductor nanowires in the presence of a strong Zeeman and Rashba spin orbit field \cite{Alicea, Lutchyn, Oreg} or a quantum dot system \cite{SauDasSarma,Fulga2013,LeijnseFlensberg}. Experiments demonstrating zero bias conductance peaks in nanowire systems \cite{Mourik,Das}, and supercurrents \cite{Sacepe2011}, Fraunhofer patterns and Shapiro steps \cite{VeldhorstJJ}, and SQUIDs \cite{Veldhorstsquid} in superconductor-topological insulator devices are steps towards the definitive detection of a Majorana bound state. More recent experiments \cite{Yazdani2014} have provided even stronger evidence for Majoranas in fabricated iron atomic chains on superconducting lead, however no evidence of braiding has yet been demonstrated.

\begin{figure}
\begin{center}
\includegraphics[width=8.5cm]{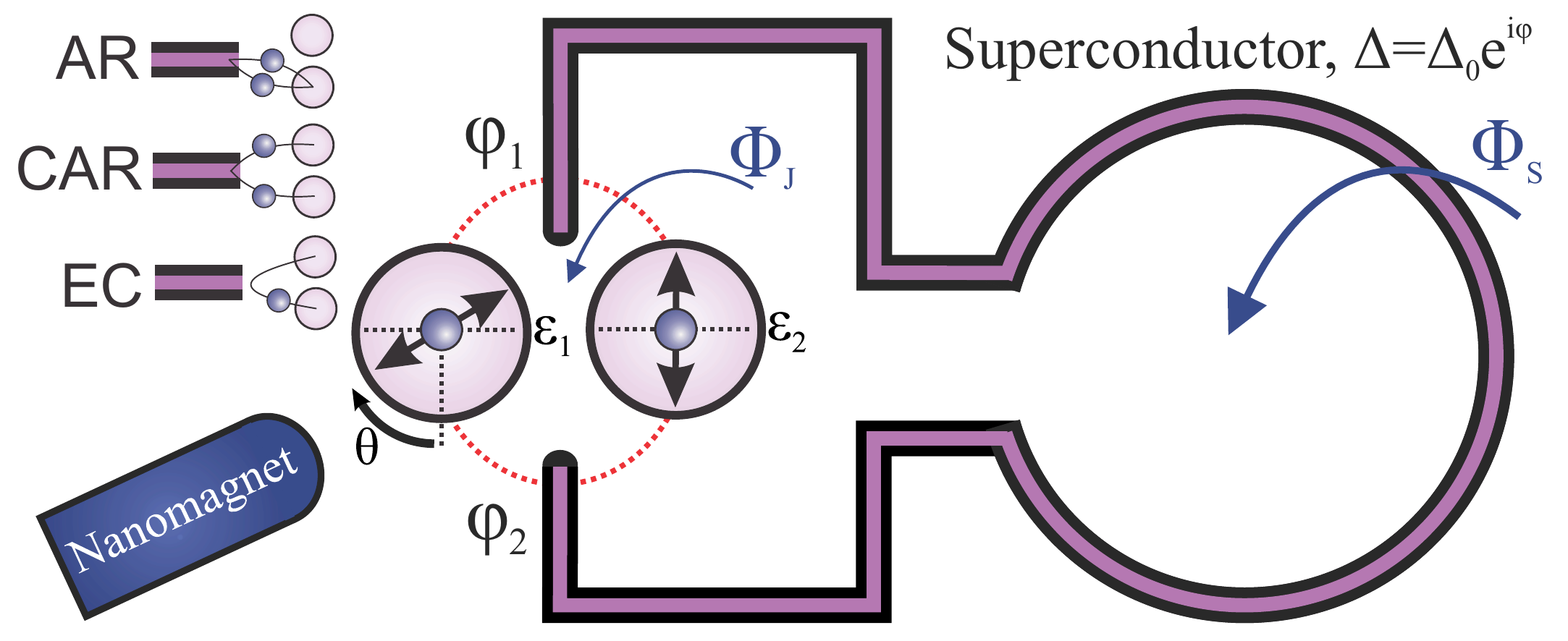}
\caption{Schematic design of the hybrid superconductor quantum dot system. A Josephson junction is formed by connecting a double quantum dot on both sides to superconducting leads. The strength of the Zeeman spin splitting in the quantum dots is determined by the external field and the stray field of the nanomagnet. We note that fields $E_Z \approx k_B T$ are sufficient for the realization of Majorana bound states. The superconducting leads form a loop, as in a RF SQUID geometry, which allows to perform a current-phase relationship measurement. The possible trajectories in the junction that determine the total supercurrent are shown in the upper left. Axes are defined in the bottom-left corner for future use.}
\label{fig:Schematic}
\end{center}
\end{figure}

Superconductor-quantum dot devices have several advantages over other proposed systems to support Majorana bound states. These systems can be lithographically defined, are strongly gate tunable and are readily operated in the few electron regime. The superconducting proximity effect is also gate tunable, which allows the stringent conditions for the presence of Majorana bound states to be satisfied. Unfortunately, Majorana bound states in the single particle formalism rely on a strong Zeeman field in combination with strong spin-orbit coupling \cite{SauDasSarma, Fulga2013} (or a strong magnetic field gradient between the dots mimicking the spin-orbit coupling \cite{LeijnseFlensberg}), making the experimental realization a serious challenge. Specifically, the Zeeman splitting biases towards the spin-polarised Kitaev state \cite{Kitaev}, while the induced $s$-wave superconducting gap $\Delta$ biases against this state. This then requires that the Zeeman splitting is large enough, so that the bias towards the Kitaev state dominates. \color{black} Recent results \cite{WolmsSternFlensberg} suggest the possibility of braiding Kramers pairs of Majorana bound states. This relaxes the requirement for time-reversal symmetry breaking, but requires protection from perturbations that mix these pairs instead.\color{black}

Fortunately, these requirements can be strongly relaxed as shown recently by Wright and Veldhorst \cite{TonyMennoPaper}, who considered the effect of strong correlations for the engineering of Majorana bound states. They demonstrated that only vanishingly small anisotropic Zeeman splitting, $\sim k_B T$, is required. The resulting Majorana bound states are described in the many-particle formalism, and although not strictly localized, they have the property of a relaxed form of localization. Specifically, the operator corresponding to a Majorana bound state is localized to a single site with respect to single creation or annihilation operator terms, but is necessarily non-local with respect to number operator terms. Since number operators do not gain a phase during braiding operations, the Majorana bound states retain a local relative phase, which is responsible for their non-Abelian behaviour. These results open new perspectives for the realization of Majorana bound states. Therefore, a rigorous derivation of how the Majorana bound states behave in transport measurements and in the presence of magnetic fields is highly valuable.

In this paper, we explore the superconductor-double quantum dot system in detail, and in particular the regions where Majorana bound states are predicted to appear. We start by considering two quantum dots proximity coupled to superconducting leads in the presence of an anisotropic magnetic field and derive an effective Hamiltonian. We investigate thoroughly the form of the excitation operators at a degeneracy point, and present concrete arguments for their being Majorana bound states. This discussion is based on the many-particle requirements for a Majorana bound state, and the existence of a continuous transition between this system and Kitaev's ground-breaking model. Finally, we investigate the Josephson supercurrent that may provide experimental evidence for Majorana bound states in these systems.

\section{\label{sec:Layout}Device layout}

Fig. \ref{fig:Schematic} shows a schematic design of the considered system. A Josephson junction is realized by connecting a double quantum dot on both sides to a superconducting lead. Two magnetic fields are present; a 'local' magnetic field produced by the stray fields of the nanomagnet together with an 'external' field from another source. Together, these determine the Zeeman splitting $E_Z$ in the dots and define the spin quantization axis, as well as determining the flux $\Phi_J$ through the quantum dot Josephson junction and the flux $\Phi_S$ through the superconducting ring. Anisotropy in the local magnetic field due to the position and strength of the nanomagnet allows for a nonzero angle $\theta$ between the spin axis of the quantum dots. We will show that Majorana bound states appear in small magnetic fields, $E_Z \approx k_B T$, and for any non-zero $\theta$. 

An important measure of the system is the supercurrent that can flow through the quantum dots via the superconducting proximity effect. This supercurrent is dependent on the flux $\Phi_J$ through the junction, and the superconducting phase difference $\phi_-=\phi_1-\phi_2$. The superconducting phase $\phi_-$ can be tuned via the flux $\Phi_S$ through the loop of the superconducting leads. This layout is similar in design to an RF SQUID consisting of a Josephson junction, and has often been used to measure the current phase relation (CPR). Since the protection of the Majorana bound states in this system is via parity conservation, this arrangement of a closed loop without macroscopic leads to the outside world should minimize quasiparticle poisoning. 

The possible trajectories of quasiparticles in the system that determine the supercurrent through the junction are shown in Fig. \ref{fig:Schematic}. A Cooper pair can tunnel from the superconducting lead to a quantum dot via Andreev reflection (AR) or split over the two quantum dots via crossed Andreev reflection (CAR). Quasiparticle tunneling between the quantum dots, via the superconducting leads, is called elastic co-tunneling (EC). Interestingly, CAR is unaffected by the magnetic field through the junction, such that the supercurrent dependence on $\Phi_J$ is determined by AR and EC. The superconducting phase difference $\phi_-$ controlled by $\Phi_S$, however, only affects AR and CAR, but not EC. We will show that these dependencies lead to novel current-phase relationships and results in a strong flexibility to realize Majorana bound states.

The requirement for only small Zeeman splitting gives several advantages over other quantum dot Majorana bound state proposals. Firstly, it opens a wider range of suitable materials. Experimentally, supercurrents through quantum dots formed in carbon nanotubes\cite{Herrero2006}, InAs nanowires \cite{Dam2006}, InAs quantum dots \cite{Buizert2007}, and graphene \cite{Dirks2011} have been observed, making them potential candidates to observed the many-particle Majorana bound states. Secondly, the experimental conditions are strongly relaxed, since the conditions where Majorana modes arise and CPR measurements are greatly simplified.

\section{\label{sec:EffHam}Derivation of effective Hamiltonian}

The Hamiltonian describing the system shown in Fig. \ref{fig:Schematic} is given by

\begin{equation}
\Hh=\Hh_S+\Hh_D+\Hh_U+\Hh_T.
\label{eqn:HamOverview}
\end{equation}

\color{black}We will describe each term in turn. The superconducting loop is modelled as a linear chain with chemical potential $\mu$, hopping strength $t_S$, and superconducting order parameter $\Delta_Se^{i\phi_j}$. Note that whilst the magnitude $\Delta_S$ is expected to be constant throughout the superconductor, the superconducting phase $\phi_j$ will change as we wind about the magnetic field $\Phi_S$. We write

\begin{align}
\Hh_S&=-t_s\sum_{\<i,j\>,\sigma}\fhatdag_{i\sigma}\fhat_{j\sigma}+\mu\sum_{j,\sigma}\fhatdag_{j\sigma}\fhat_{j\sigma}\nonumber\\
&\;\;\;\;\;+\sum_{j}(\Delta_S e^{i\phi_{j}}\fhatdag_{j\+}\fhatdag_{j\-}+\text{h.c.}).
\end{align}

Here, $\<i,j\>$ denotes pairs of nearest neighbors. We will ultimately only be concerned with the phases on the ends of the superconductors - let us label these $\phi_1$ and $\phi_N$, and $\phi_{\pm}=\phi_1\pm\phi_N$. Importantly, the phase difference $\phi_-$ can be tuned to a high degree of accuracy by adjusting $\Phi_S$.

\color{black}

The dots ($\Hh_D$) have an on-site potential $\epsilon_j$, and are considered in the presence of a small (but non-zero) magnetic field, leading to a Zeeman energy splitting $E_Z$

\begin{equation}
\Hh_D=\sum_{j=1}^2\sum_{\sigma}\epsilon_j\chatdag_{j\sigma}\chat_{j\sigma}-E_Z\sum_{j}(\chatdag_{j\+}\chat_{j\+}-\chatdag_{j\-}\chat_{j\-}).
\end{equation}
We consider the quantum dots to be operated in the few electron regime, where a strong Coulombic repulsion is present between two electrons on the same dot. We model this as a Hubbard-style interaction

\begin{equation}
\Hh_U=U\sum_{j=1}^2\chatdag_{j\+}\chat_{j\+}\chatdag_{j\-}\chat_{j\-}.
\label{eqn:HamU}
\end{equation}

The two superconductors are coupled to the dots via the proximity effect, which allows for electron tunneling between either \color{black}end of the\color{black} superconductor and either dot:

\color{black}
\begin{equation}
\Hh_T=\sum_{n=\{1,N\}}\sum_{j=\{1,2\}}\sum_{\sigma}\Gamma_{n,j}(\fhatdag_{n\sigma}\chat_{j\sigma}+\text{h.c.}).
\label{eqn:HamT}
\end{equation}
\color{black}
We assume that the tunnel coupling $\Gamma_{n,j}$ of the two dots have the same amplitude $\Gamma$, and study the effect of finite phase difference. In practice this phase difference is realized by the magnetic flux through the quantum dot Josephson junction, as shown in Fig. \ref{fig:Schematic}. In the Peierls substitution, the tunneling obtains a phase proportional to the size of the field

\begin{equation}
\Gamma_{n,j}=\Gamma\exp\left(-i\frac{\pi}{\Phi_0}\int_j^n\mathbf{A}\cdot d\mathbf{r}\right).
\end{equation}

Here, $\Phi_0=h/2e$ is the magnetic flux quantum, and we integrate along lines between the superconductors and the dots (see Fig. \ref{fig:Schematic}). This is a slight simplification, as the electrons do not strictly travel along any given line, but the result is essentially the same. Choosing the gauge $\mathbf{A}=-By\hat{x}$ (refer to Fig. \ref{fig:Schematic} for axis), we calculate

\begin{equation}
\Gamma_{n,j}=\Gamma\exp\left(\pm i \pi \frac{\Phi_J}{\Phi_0}\right),
\end{equation}

where $\Phi_J$ is the enclosed flux, and the positive sign is taken when the path travels anticlockwise about the origin.

As all terms in the Hamiltonian involving the superconductors are quadratic, they may be removed to write down an effective Hamiltonian for the dots via an integration over Grassman variables \cite{AltlandSimons}. To do this, we write down the partition function of the system

\begin{align}
\Zz&=\int D[\bar{\Psi},\Psi]\exp(-S[\bar{\Psi},\Psi]),\label{PartitionFunction}\\
S[\bar{\Psi},\Psi]&=\int_0^{\beta}d\tau\left[\bar{\Psi}\partial_{\tau}\Psi+\Hh[\bar{\Psi},\Psi]\right],
\end{align}

where $\Hh$ is the functional form of the Hamiltonian. Here, $\Psi$ and $\bar{\Psi}$ are vectors of Grassman variables, which are the eigenvalues of the annihilation operators for some Fermionic coherent state $|\psi\>$ \cite{AltlandSimons}. We use the following notation to separate the Grassman variables associated with the dots from those associated with the superconductor,

\color{black}
\begin{equation}
\chat_{i\sigma}|\psi\>=\psi_{i\sigma}|\psi\>,\;\;\fhat_{j\sigma}|\psi\>=\phi_{j\sigma}|\psi\>.
\end{equation}
\color{black}

Then we can define

\begin{equation}
\Psi=\left(\begin{array}{c}\psi\\\phi\end{array}\right),\;\;\bar{\Psi}=\left(\begin{array}{cc}\bar{\psi}&\bar{\phi}\end{array}\right).
\end{equation}

\color{black}
Here $\psi$($\phi$) contains the $\psi_{i\sigma}$($\phi_{j\sigma}$) terms, and $\bar{\psi}$ ($\bar{\phi}$) contains their adjoints, which are defined by $\<\psi|\chatdag_{i\sigma}=\<\psi|\bar{\psi}_{i\sigma}$ and $\<\psi|\fhatdag_{j\sigma}=\<\psi|\bar{\phi}_{j\sigma}$. As we are dealing with a superconducting system, it is necessary to use an electron-hole Nambu basis, and include the adjoint variables $\bar{\psi}_{i\sigma}$ ($\bar{\phi}_{j\sigma}$) in the $\psi$ ($\phi$) vector (which doubles in size). We can then expand $\Hh[\bar{\Psi},\Psi]$ into the terms from equation \ref{eqn:HamOverview}, replacing $\Psi$ with either $\psi$ or $\phi$ depending on which species is being considered. Furthermore, quadratic Hamiltonian terms can be rewritten as matrix products, for example $\Hh_D[\bar{\psi},\psi]\rightarrow \bar{\psi}H_D\psi$. In this notation, our action becomes
\color{black}
\begin{align}
S[\bar{\Psi},\Psi]&=\int_0^{\beta}d\tau\left[\bar{\psi}\partial_{\tau}\psi+\Hh_U[\bar{\psi},\psi]+\bar{\psi}H_D\psi\right.\nonumber\\&\left.+\bar{\phi}\Gg^{-1}\phi+\bar{\psi}M\phi+\bar{\phi}M^{\dag}\psi\right].
\label{eqn:EffAction}
\end{align}

Here, we have defined $\Gg^{-1}=\partial_{\tau}+H_S$, and split the terms from $\Hh_T$ into $M$, which contains the information on tunneling from the superconductors to the dot, and its adjoint $M^{\dag}$. Note that we are treating each Grassman variable as independent from its corresponding adjoint.  We now shift all $\phi$ dependence to a single term by completing the square

\begin{align}
S[\bar{\Psi},\Psi]&=\int_0^{\beta}d\tau\left[\bar{\psi}\partial_{\tau}\psi+\Hh_U[\bar{\psi},\psi]+\bar{\psi}H_D\psi\right.\nonumber\\
&\left.+(\bar{\phi}\Gg^{-1}+\bar{\psi}M)\Gg(\Gg^{-1}\phi+ M^{\dag}\psi)-\bar{\psi}M\Gg M^{\dag}\psi\right].
\end{align}

The term containing the $\phi$ dependence may be integrated out to give a constant \cite{AltlandSimons}. Our effective Hamiltonian then consists of the original $\Hh_D$ and $\Hh_U$ terms, and a new term which was the remainder from completing the square

\begin{equation}
\Hh_{new}[\psi,\bar{\psi}]=-\bar{\psi}M\Gg M^{\dag}\psi.
\end{equation}

\begin{figure*}
\begin{tabular}{cc}
\includegraphics[width=8.8cm]{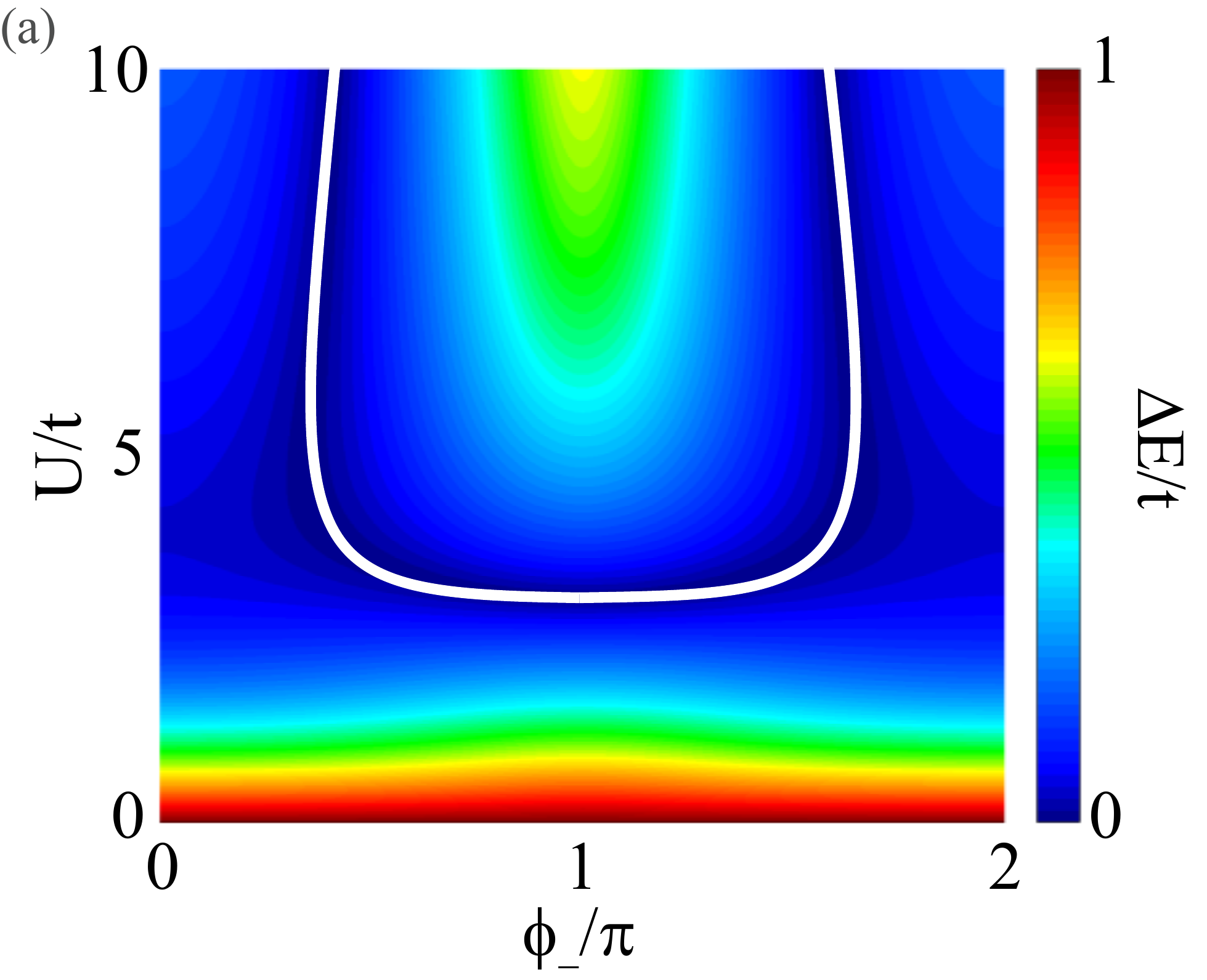}&
\includegraphics[width=8.8cm]{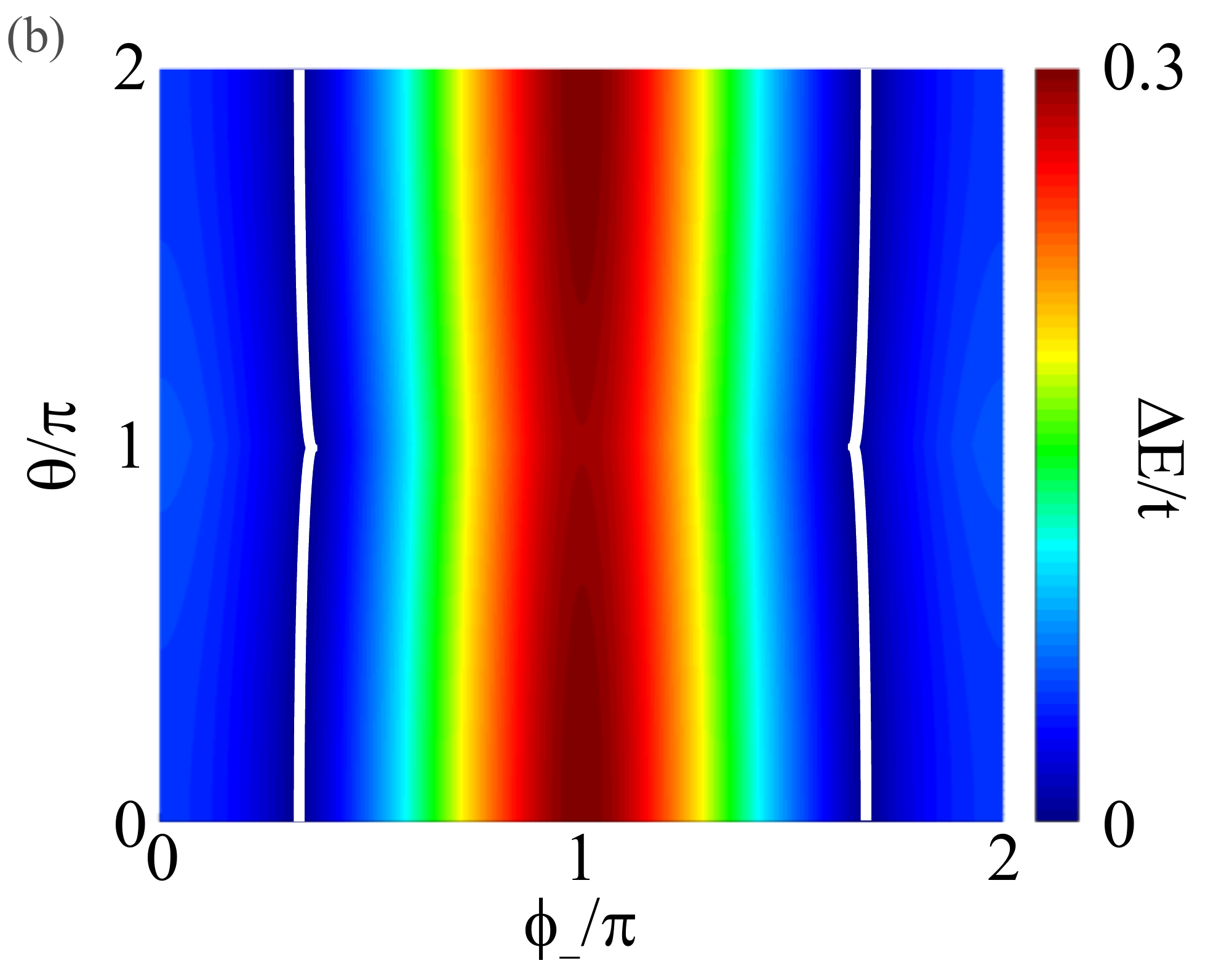}\\
\includegraphics[width=8.8cm]{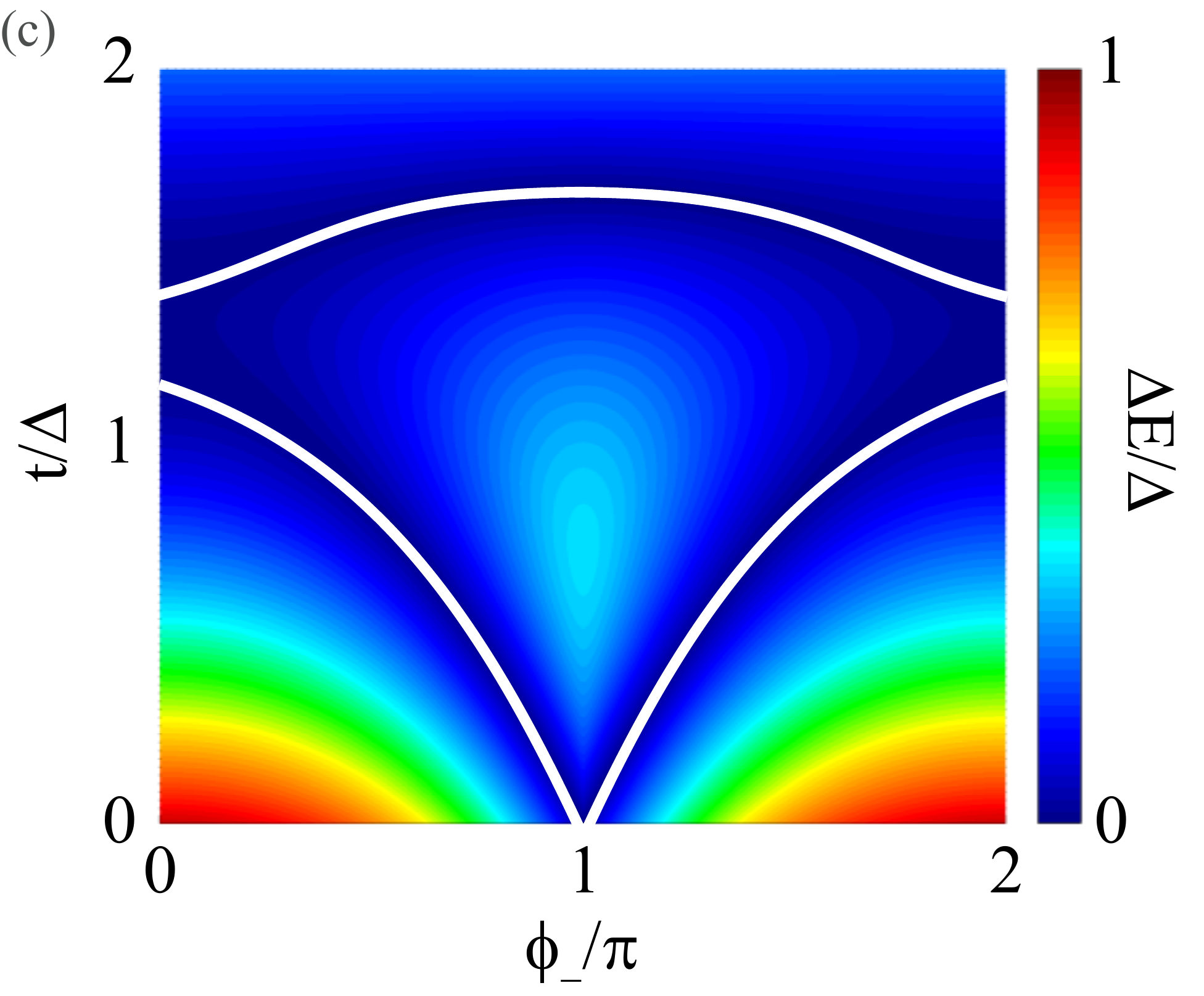}&
\includegraphics[width=8.8cm]{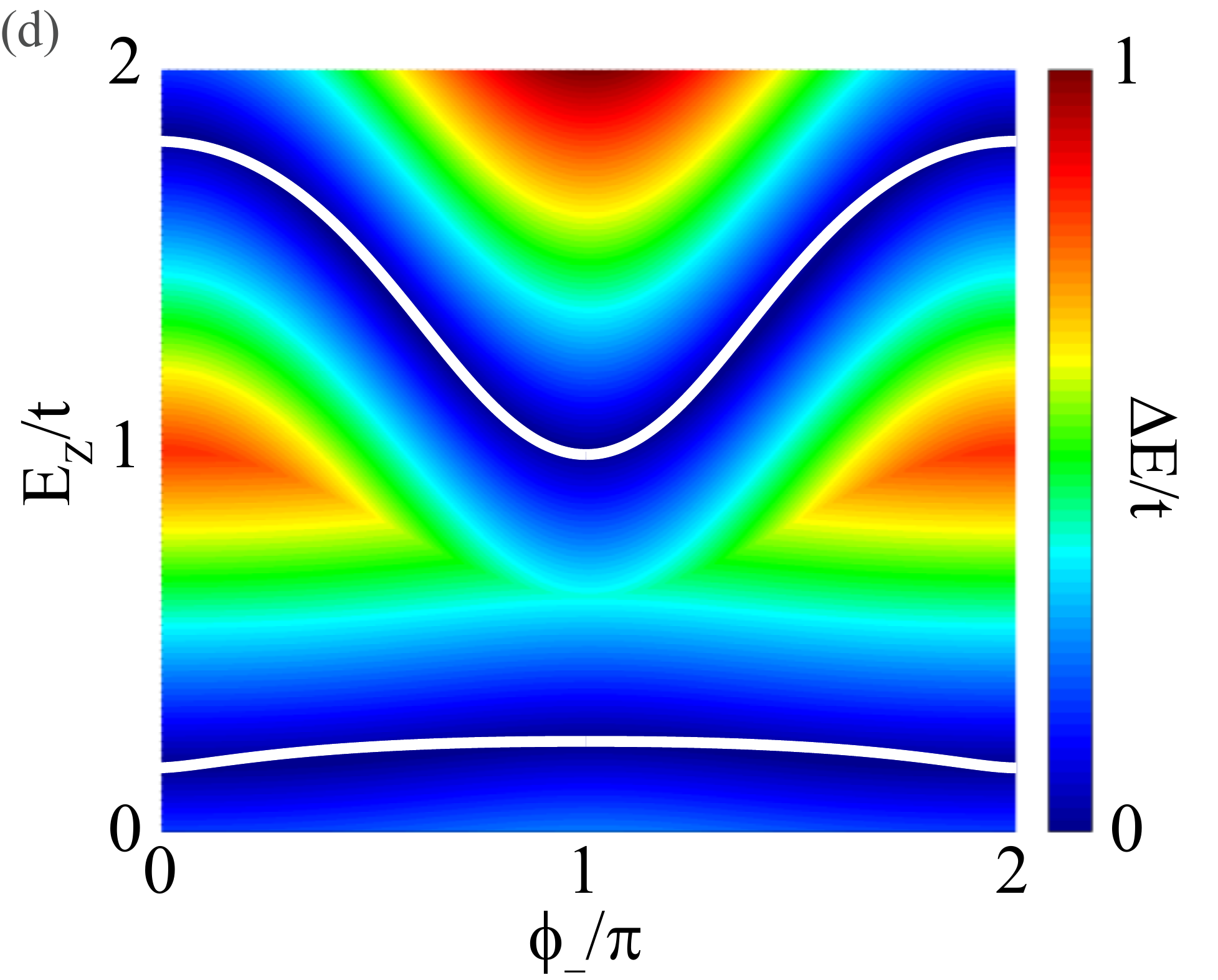}
\end{tabular}
\caption{Surface plots of the even-odd excitation energy for various pairs of parameters in the double quantum dot. All include the difference in the experimentally tunable superconducting angle $\phi_-$. Of interest are points where this energy is near \ zero (white lines), as at these points we expect to find Majorana bound states. \color{black} As our effective Hamiltonian is an approximation, we should find corrections to the exact position of these lines at higher orders in our cumulant expansion (Eq.\ref{eqn:CumulantExpansion}), but whilst these remain small, they will not break the degeneracy as they do not mix the even and odd sectors. (a): increasing the Coulomb repulsion $U$ gives a critical value above which a degenerate state is achieved for certain values of $\phi_-$. This critical $U$ value is dependent on the other parameters, here $t=\Delta$, $E_Z=0.001t$, $\theta=0$, and then the critical value is approximately $U=3t$ (this is exact when $E_Z=0$). (b): changing the spin-orbit coupling when $E_Z\approx 0$ has minimal impact on the ground state. When $E_Z=0$, the spin rotation is simply a basis change, and does not affect the eigenstate energies. (c): there is a wide range of $t/\Delta$ for which a degenerate ground state can occur (here at $U=5$, $\theta=0$, $E_Z=0.001$). \color{black} As $U\rightarrow 0$, the range of degeneracy shrinks, requiring a smaller and smaller effective hopping. For large $U$, the higher energy $t/\Delta$ degeneracy breaks away and shifts towards $t/\Delta=\infty$, and there is always some range of $t/\Delta$ where the degeneracy cannot be found. (d): increasing the strength of the Zeeman splitting $E_Z$ can also take the system to a degeneracy. Here, $t=\Delta=U/2$.}
\label{fig:CharacterPlots}
\end{figure*}

This term can be evaluated via matrix multiplication, but we must first specify a basis, which we do by explicitly writing down the vectors $\psi$ and $\phi$. The coupling Hamiltonian $\Hh_T$ only includes operators from the $N$th site of either superconductor, and thus non-zero contributions will come only from products of matrix elements from these sites. Equivalently, for our purposes we can use  a basis for $\phi$ which only includes the $N$th site terms, reducing it to a manageable size. We write

\color{black}
\begin{align}
\bar{\psi}&=\left(\bar{\psi}_{1\+},\bar{\psi}_{2\+},\bar{\psi}_{2\-},\bar{\psi}_{1\-},\psi_{1\+},\psi_{2\+},\psi_{2\-},\psi_{1\-}\right),\\
\bar{\phi}&=\left(\bar{\phi}_{1\+},\bar{\phi}_{N\+},\bar{\phi}_{N\-},\bar{\phi}_{1\-},\phi_{1\+},\phi_{N\+},\phi_{N\-},\phi_{1\-}\right).
\end{align}
\color{black}

We assume for now that no flux passes between the dots ($\Gamma_{n,j}=\Gamma$), and then our coupling matrix $M$ is

\begin{align}
M=M^{\dag}&=\Gamma\left(\begin{array}{cc}M_{AA}&0\\0&-M_{AA}\end{array}\right),\\
M_{AA}&=\left(\begin{array}{cccc}1&1&0&0\\1&1&0&0\\0&0&1&1\\0&0&1&1\end{array}\right).\nonumber
\end{align}

\color{black}
To calculate $\Gg$, we use the Matsubara representation. We make the following substitution
\begin{equation}
\psi(\tau)=\frac{1}{\sqrt{\beta}}\sum_{\omega_n}\psi_ne^{-i\omega_n\tau},
\end{equation} 
where $\omega_n$ are the Matsubara frequencies $\omega_n=(2n+1)\frac{\pi}{\beta}$. Then, we have $\partial_{\tau}\rightarrow -i\omega_n$, and we can calculate
\color{black}

\begin{align}
\Gg^{-1}&=\left(\begin{array}{cc}G_{AA}&G_{AB}\\G_{AB}^{\dag}&G_{BB}\end{array}\right),\nonumber\\&G_{AA}=(\mu-i\omega_n)I,\;\;G_{BB}=-(\mu+i\omega_n)I,\nonumber\\
G_{AB}&=\Delta_S\left(\begin{array}{cccc}0&0&0&e^{i\phi_1}\\0&0&e^{i\phi_2}&0\\0&-e^{i\phi_2}&0&0\\-e^{i\phi_1}&0&0&0\end{array}\right).\nonumber
\end{align}

The inverse to this matrix is

\begin{equation}
\Gg=\frac{1}{\mu^2+\omega_n^2+\Delta_S^2}\left(\begin{array}{cc}-G_{BB}&G_{AB}\\G_{AB}^{\dag}&-G_{AA}\end{array}\right),
\end{equation}

and from this we calculate

\begin{align*}
&-M\Gg M^{\dag}=\frac{-\Gamma^2}{\mu^2+\omega_n^2+\Delta_S^2}\\&\times\left(\begin{array}{cc}M_{AA}(i\omega_n+\mu)M_{AA}&-M_{AA}G_{AB}M_{AA}\\-M_{AA}G_{AB}^{\dag}M_{AA}&M_{AA}(i\omega_n-\mu)M_{AA}\end{array}\right),\\
&M_{AA}^2=2M_{AA},\\&M_{AA}G_{AB}M_{AA}=2(e^{i\phi_1}+e^{i\phi_2})\left(\begin{array}{cccc}0&0&1&1\\0&0&1&1\\-1&-1&0&0\\-1&-1&0&0\end{array}\right).
\end{align*}

\color{black}
It remains to invert the Fourier transform, but this will return an action which is non-local in time
\begin{align}
&-\frac{1}{\beta}\sum_n\bar{\psi}_n M\Gg M^{\dag}\psi_n\nonumber\\
&\;\;\;\;\;=-\frac{1}{\beta^2}\int_0^\beta d\tau d\tau'\sum_n\bar{\psi}(\tau)\left(M\Gg M^{\dag}\right)_n\psi(\tau')e^{i\omega_n(\tau'-\tau)}.
\end{align}

Only one summation over $n$ needs to be computed, and this can be performed using complex integration techniques \cite{AltlandSimons}:
\begin{align}
&\sum_n\frac{1}{\mu^2+\omega_n^2+\Delta_S^2}e^{i\omega_n(\tau-\tau')}\\
&\;\;\;\;\;=\frac{\beta}{e^{-K\beta}+1}\frac{1}{K}e^{-K|\tau-\tau'|},
\end{align}
where $K=\sqrt{\Delta_S^2+\mu^2}$. We then approximate this by the following function
\begin{equation}
\frac{\beta}{e^{-K\beta}+1}\frac{1}{K^2}\delta(\tau-\tau').
\end{equation}

This allows us to write the action as 
\begin{widetext}
\begin{align}
S[\bar{\psi},\psi]&=S_{eff}[\bar{\psi},\psi]+S_{pert}[\bar{\psi},\psi]\\
S_{eff}[\bar{\psi},\psi]&=\int_0^\beta d\tau\left[\bar{\psi}\partial_\tau\psi+\Hh_U[\bar{\psi},\psi]+\bar{\psi} H_D\psi-\bar{\psi}\frac{\Gamma^2}{K^2(e^{-K\beta}+1)}\Mm\psi\right]\\
S_{pert}[\bar{\psi},\psi]&=\int_0^{\beta}d\tau d\tau'\bar{\psi}\left(\frac{\Gamma^2}{K^2(e^{-K\beta}+1)}\delta(\tau-\tau')-\frac{\Gamma^2}{K(e^{-K\beta}+1)}e^{-K|\tau-\tau'|}\right)\Mm\psi\label{eqn:SPert}\\
\Mm&=\left(\begin{array}{cc}M_{AA}(-K+\mu)M_{AA}&-M_{AA}G_{AB}M_{AA}\\-M_{AA}G_{AB}^{\dag}M_{AA}&M_{AA}(-K-\mu)M_{AA}\end{array}\right)
\end{align}
\end{widetext}

Our effective action $\Ss_{eff}$ is local in time, and so we can extract an effective Hamiltonian by undoing the procedure used to write equation \ref{eqn:EffAction}.\color{black} We obtain an elastic co-tunneling (EC) term

\begin{equation}
\Hh_{ct}=t\sum_{\sigma}\chatdag_{1,\sigma}\chat_{2,\sigma} + \text{h.c,}
\end{equation}

and crossed (CAR) and normal (AR) Andreev reflection terms

\begin{equation}
\Hh_{Ar}=\Delta e^{i\phi_+/2}\cos(\phi_-/2)\sum_{i,j}\chatdag_{i\+}\chatdag_{j\-}+\text{h.c,}
\end{equation}

\color{black}
where $\phi_{\pm}=\frac{1}{2}(\phi_1\pm\phi_2)$. $t$ and $\Delta$ can be read immediately from the effective action
\begin{align}
t=\frac{-2\Gamma^2\mu}{K^2(e^{-K\beta}+1)}\\
\Delta=\frac{-2\Delta_S\Gamma^2}{K^2(e^{-K\beta}+1)}.
\end{align}

Corrections to the effective action can be calculated via a cumulant expansion \cite{MengPaper,RozhkovArovas}. To do this, we write
\begin{equation}
\Zz\approx\int\Dd[\bar{\psi},\psi]e^{-S_{eff}}(1-S_{pert}+\frac{1}{2}S_{pert}^2-\ldots).
\label{eqn:CumulantExpansion}
\end{equation}
The first order correction is the thermal expectation value of $S_{pert}$ in our effective approximation. This can be written
\begin{equation}
\<S_{pert}\>_{eff}=\int_0^{\beta}d\tau \Gg_{pert}(\tau)_{i,j}\<\Tt_{\tau}\bar{\psi}_i(\tau)\psi_j(0)\>_{eff}.
\end{equation}
Here, $\Gg_{pert}$ is taken from the effective action (Eq.\ref{eqn:SPert}) by writing $S_{pert}[\bar{\psi},\psi]=\int)^{\beta}d\tau d\tau'\bar{\psi}\Gg_{pert}(\tau-\tau')\psi$. Exact calculations of $\<\Tt_{\tau}\bar{\psi}_i(\tau)\psi_j(0)\>_{eff}$ are done in \cite{MengPaper} for a similar system to ours, and are of the order of $e^{-|\Delta E|}$, where $\Delta E$ is the spacing between lowest energy levels. As we are interested in our system at degeneracy, this will be $\approx 1$. The integral over $\tau$ can be calculated by substituting in $\Gg_{pert}$
\begin{equation}
\int_0^\beta d\tau\Gg_{pert}=\frac{\Gamma^2}{K^2(e^{-K\beta}+1)}e^{-K\beta}\Mm.
\end{equation}
This becomes exponentially small in the large $\beta$ limit, justifying our approximation at low temperatures. Higher order terms in our cumulant expansion will scale at higher powers of $e^{-K\beta}$. This is thus an acceptable approximation at low temperatures, but we need to be aware of two sources of error that come from this approximation. Firstly, it will cause corrections to the relative energy levels (as in \cite{MengPaper}), and secondly it will give any quasiparticles a finite lifetime on the order of $\hbar/\<S_{pert}\>_{eff}$. The energy corrections will not mix the even and odd parity sections of the effective Hamiltonian, and as we will show in the next section, this implies they do not prevent the existence of Majorana bound states. However, the finite quasiparticle lifetime will need to be accounted for in any experimental design.
\color{black}

Previous results in the literature show that elastic co-tunneling is to lowest order in tunneling amplitude equal in magnitude and opposite in sign to crossed Andreev reflection\cite{Falci2001}. When higher order terms are included, EC has a larger contribution \cite{Kalenkov2007}, however the electromagnetic environment \cite{Yeyati2007} and Coulomb interactions \cite{Recher2001} can result in CAR being dominant instead. This is demonstrated in recent experiments \cite{Hofstetter2009, Herrmann2010}.

An effective spin-orbit coupling is obtained by rotating the local magnetic field on dot $2$, using the nanomagnet shown in Fig. \ref{fig:Schematic}. This can be treated as a uniform spin rotation on the respective site, sending $\chat^{(\dag)}_{2\sigma}\rightarrow\cos(\theta/2)\chat^{(\dag)}_{2\sigma}+\sigma\sin(\theta/2)\chat^{(\dag)}_{2\bar{\sigma}}$. The effective Hamiltonian is then \cite{TonyMennoPaper}:

\begin{widetext}
\begin{align}
\Hh_{eff}&=\sum_{j,\sigma}\epsilon'_j\chatdag_{j,\sigma}\chat_{j,\sigma}-E_Z\sum_{j}(\chatdag_{j\+}\chat_{j\+}-\chatdag_{j\-}\chat_{j\-})+U\sum_j\chatdag_{j\+}\chat_{j\+}\chatdag_{j\-}\chat_{j\-}+t\sum_{\sigma}\left(\cos(\theta/2)\chatdag_{1,\sigma}\chat_{2,\sigma}+\sigma\sin(\theta/2)\chatdag_{1,\sigma}\chat_{2,\bar{\sigma}}+\text{h.c.}\right)\nonumber\\&+\Delta e^{i\phi_+/2}\cos(\phi_-/2)\left(\sum_i\chatdag_{i\+}\chatdag_{i\-}+\sum_{\sigma}(\sigma\cos(\theta/2)\chatdag_{1\sigma}\chatdag_{2\bar{\sigma}}-\sin(\theta/2)\chatdag_{1\sigma}\chatdag_{2\sigma}+\text{h.c.})\right).
\label{EffectiveHamiltonian}
\end{align}
\end{widetext}

The effect of the magnetic flux through the quantum dot Josephson junction can be found by altering $\Gamma_{n,j}$ for the coupling matrix $M$. This results in the elastic co-tunneling term being universally multiplied by a factor that comes from interference between the two possible paths from one dot to the other.

\begin{equation}
t\rightarrow t\cos\left(\frac{\pi}{2}\frac{\phi_J}{\phi_0}\right),
\end{equation}

Crossed Andreev reflection is unaffected by the flux through the junction, as the electron-hole time reversed partners cancel the magnetic field. However, for normal Andreev reflection, this new source of interference sums with the phase difference between the possible parent superconductors. This changes

\begin{equation}
\cos(\phi_-/2)\chatdag_{i\+}\chatdag_{i\-}\rightarrow\cos\left(\phi_-/2\pm \frac{\pi}{2}\frac{\phi_J}{\phi_0}\right)\chatdag_{i\+}\chatdag_{i\-},
\end{equation}

where the positive sign is taken for $i=1$ and negative sign for $i=2$. 

\section{\label{sec:Characterization}Degenerate ground states}
For this section we will assume that the effective on-site potential of the two dots has been tuned to the chemical potential of the superconductors, which we define as our zero of energy ($\epsilon_1+t=\epsilon_2+t=0$). A discussion of the effects of the on-site energies deviating from this `sweet spot' has been presented elsewhere \cite{LeijnseFlensberg}. We will assume for this section as well that there is no magnetic flux passing between the dots ($\Lambda=0$).

To investigate the appearance of Majorana bound states in this system, we first investigate what freedom we have to tune to a ground state degeneracy. Here, we explicitly require this degeneracy to be between states with different particle number parity, as these are protected against mixing. Due to this protection, we can separate our basis states into even and odd sectors, reducing the Hamiltonian to an $8\times 8$ matrix for each. These sectors cannot be split further by particle number or spin when the superconducting or anisotropic Zeeman terms are respectively present.

To characterise the system, in Fig. \ref{fig:CharacterPlots} we present surface plots of the difference in energy between the lowest energy even and odd particle number eigenstates (which we call the even-odd excitation energy), for various sets of parameters. For fixed values of $t$ and $\Delta$, there exists a minimum value of $U$ required for the degeneracy we require (at $t=\Delta$ and $E_Z=0$, this is at $U=3t$; see Fig. \ref{fig:CharacterPlots}.a). \color{black} It should be noted that there exists no degeneracy at $U=0$, $Z=0$ for any $t$ or $\Delta$; crossed Andreev reflection prevents our system realising a spinful Kitaev chain in the non-interacting limit. \color{black}

\color{black}
Above the minimum $U$ value, the degeneracy can always be realized by tuning the superconducting phase difference $\phi_-$. This is important, as whilst other parameters will be relatively constrained in an experiment, the relative superconducting phases are freely tunable via changing the flux $\Phi_S$ through the superconducting loop, as shown in Fig. \ref{fig:Schematic}. Furthermore, higher order corrections from the cumulant expansion in Sec.\ref{sec:EffHam} will not mix the even and odd sectors, and thus, though they might cause corrections to the position of the lines of degeneracy, they will not remove the ability to tune to them.

It is also important to note that the even-odd degeneracy can be reached for $E_Z$ at and near $0$. When $E_Z\approx \pm k_BT$, we expect a single MBS to be present on either dot; the MBS will be different depending on the sign of $E_Z$. When $E_Z=0$, the well-known Kramers degeneracy is also present in the odd parity states. The two species of MBS on each dot then should form Kramers pairs, as outlined in \cite{WolmsSternFlensberg}. However, our system as described does not have the means for protecting against the mixing of the Kramers pairs, and so gapping out one of the species by a small magnetic field  $E_Z>k_B T$ is preferable. As we are requiring only small $E_Z$, the spin-orbit coupling angle $\theta$ will have negligible impact on the eigenenergies (see Fig.\ref{fig:CharacterPlots}.b).
\color{black}

In the limit that $U\rightarrow\infty$, the system can no longer support states containing two electrons on a single dot. This reduces our basis to five even particle number and four odd particle number states. We break the Kramers degeneracy and diagonalize our Hamiltonian for non-zero $E_Z$, but then consider the form of the wavefunctions as $E_Z\rightarrow 0$. This provides a good approximation for the negligible $E_Z$ case, where the Kramers degeneracy is only broken on the order of the temperature.\color{black} We will detail the $E_Z>0$ results here - our procedure easily generalises to the $E_Z<0$ sector, and to the $E_Z=0$ sector with the mixing of the Kramers pairs. \color{black}

At \color{black} $E_Z=0^+$ \color{black}, we find an even particle number state $|\Psi_E\>$ with energy $\epsilon=-\sqrt{2}\Delta|\cos(\phi_-)|$, and an odd particle number state $|\Psi_O\>$ with energy $\epsilon=-t$. The requirement for a degenerate ground state then is that $\sqrt{2}\Delta|\cos(\phi_-)|=t$. This requirement will be satisfied at some $\phi_-$ whenever $\sqrt{2}\Delta > t$. From Fig. \ref{fig:CharacterPlots}, we see that this upper bound is lowered if $U$ is finite. When $\cos(\phi_-)>0$, the form of the two lowest energy eigenstates are

\begin{align}
|\Psi_E\>&=\frac{1}{2}\left[\sqrt{2}e^{i\phi_+/4}-e^{-i\phi_+/4}\cos(\theta/2)(\chatdag_{1\+}\chatdag_{2\-}+\chatdag_{2\+}\chatdag_{1\-})\right.\nonumber\\&\left.\hspace{0.8cm}+e^{-i\phi_+/4}\sin(\theta/2)(\chatdag_{1\+}\chatdag_{2\+}+\chatdag_{1\-}\chatdag_{2\-})\right]|v\>,\\
|\Psi_O\>&=\frac{1}{\sqrt{2}}\left[\cos(\theta/4)(\chatdag_{2\+}-\chatdag_{1\+})\right.\nonumber\\&\left.\hspace{2.5cm}+\sin(\theta/4)(\chatdag_{2\-}+\chatdag_{1\-})\right]|v\>.
\end{align}

\section{\label{sec:Evidence for Majorana bound states} Evidence for Majorana bound states}
\begin{figure}
\begin{center}
\includegraphics[width=8.5cm]{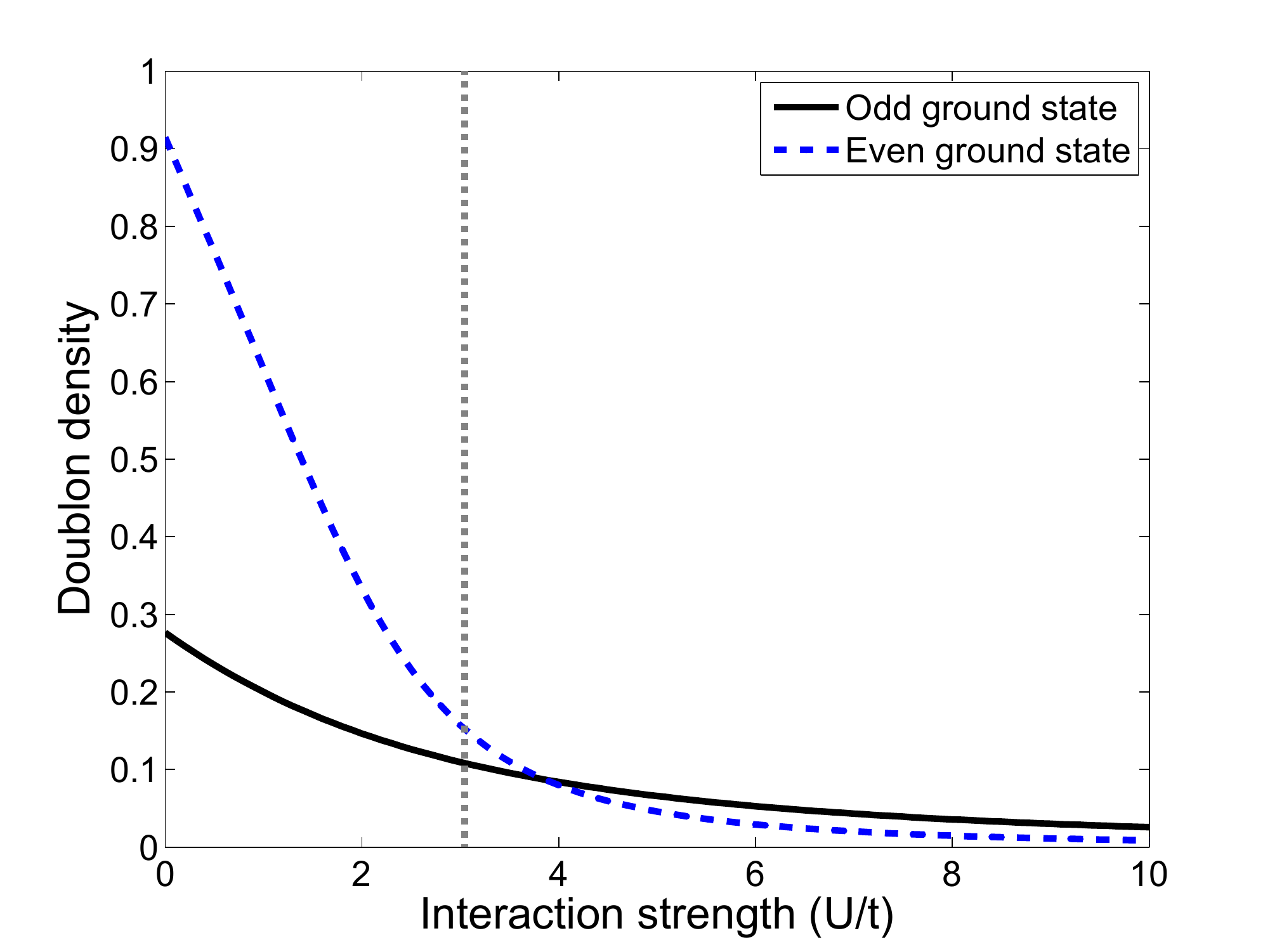}
\caption{Doublon (doubly occupied site) density for the lowest energy even and odd states of the double quantum dot as $U$ increases. Other parameters are $\Delta=t$, $\theta=\phi_-=\phi_+=0$, $E_Z=0$. We see that the doublon density is fairly small for most $U>3t$ (to the right of the dotted line), which is where the ground state degeneracy is present. Thus the infinite $U$ approximation should be reasonably accurate for most systems where a degenerate ground state is present.}
\label{fig:DoublonDensity}
\end{center}
\end{figure}

We now present evidence for the existence of Majorana bound states when our system is close to a degeneracy. We continue to take the $U\rightarrow\infty$ limit, which is equivalent to an assumption that doublons are not present. In Fig. \ref{fig:DoublonDensity}, we see that for $U$ much greater than the critical value the doublon density has dropped to a negligible amount. This implies that the following arguments should hold for a large range of finite $U$ also. Experimentally, quantum dots can have charging energies of several meV, so that $U$ will typically be large.


As in any interacting problem, the form of the excitations between two states is difficult to determine. It is possible to show for this system that no excitation operators between $|\Psi_E\>$ and $|\Psi_O\>$ may consist only of single creation or annihilation operators. For, consider the part of the operator that would excite $|\Psi_O\>$ to the two particle basis state components of $|\Psi_E\>$. This must not generate doublons under our assumption, and so it must take the form

\begin{align*}
e^{-i\phi_+/4}A\left[-\frac{\cos(\theta/2)}{\cos(\theta/4)}\chatdag_{2\-}-\frac{\cos(\theta/2)}{\sin(\theta/4)}\chatdag_{2\+}\right]\\
+e^{-i\phi_+/4}B\left[\frac{\cos(\theta/2)}{\cos(\theta/4)}\chatdag_{1\-}-\frac{\cos(\theta/2)}{\sin(\theta/4)}\chatdag_{1\+}\right].
\end{align*}

Then, evaluating the action of our excitations upon $|\Psi_O\>$, we find the following two equations which needs to be fulfilled:
\begin{align*}
(A+B)\cos(\theta/2)\tan(\theta/4)&=\frac{1}{\sqrt{2}}\sin(\theta/2)\\
(A+B)\cos(\theta/2)\cot(\theta/4)&=-\frac{1}{\sqrt{2}}\sin(\theta/2).
\end{align*} 
However, these can never be satisfied simultaneously! As such, an excitation made out of single products of creation and annihilation operators  is not possible, which forces our solutions to differ significantly from those studied in \cite{Alicea}.

\color{black}
\begin{figure}
\includegraphics[width=8.5cm]{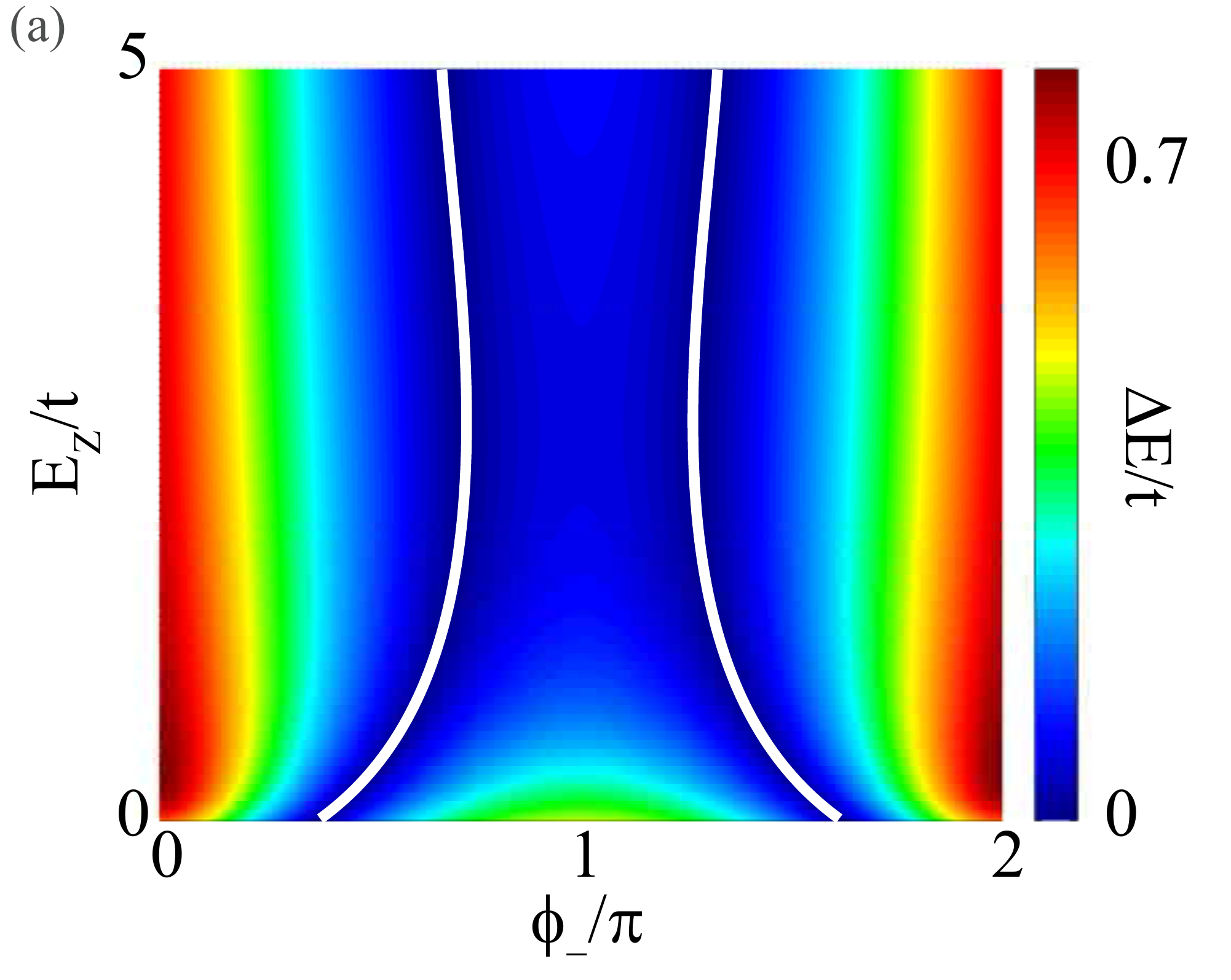}
\caption{Surface plot of the even-odd excitation energy showing the spinless Kitaev phase at large $\epsilon_1=\epsilon_2=E_Z$, and the DQD phase at vanishing $E_Z$. Our other parameters are $\Delta=\sqrt{2}t$, $\theta=\pi/2$, and $U=10$. We see that we can continuously transition from a system supporting Kitaev Majorana bound states (which are achieved as $E_Z\rightarrow\infty$ and $\Delta\cos(\phi_-/2)=t$) to the Majorana bound states we study in this paper (where $E_Z/t\approx k_B T$), whilst maintaining a degeneracy (white lines) at all times. This gives further evidence that our description of the system as containing Majorana bound states is accurate.}
\label{fig:ChangingPhase}
\end{figure}
\color{black}

This result is not surprising. In general only non-interacting systems can be guaranteed to have excitations consisting of single products of creation and annihilation operators. However, without this restriction, there is a large degree of freedom in our choice of possible operators that excite between our ground states. The question then is; if we were to write down an operator that has the form of a Majorana bound state, would we be correct in doing so?

The properties of Majorana bound states in non-interacting systems are well-known \cite{SimonReview,AliceaReview,Kitaev}. We wish to interpret the properties of these non-interacting excitations in terms of the states they excite between. Then, extending to the interacting case, if our system has these properties the Majorana picture will be the correct way to view it. 

Firstly, to have a zero energy excitation, we require $|\Psi_O\>$ and $|\Psi_E\>$ to be degenerate, as has been discussed in the previous section. Then, we note that a general non-interacting Majorana can be written $\gamma=\hat{C}+\hat{C}^{\dag}$, where $\hat{C}^{\dag}$ is a sum of creation operators, and $\hat{C}$ is a sum of annihilation operators. We want both of these terms to excite between the ground states, as otherwise $\hat{C}$ and $\hat{C}^{\dag}$ are the excitations themselves, not $\gamma$. As such, we require \emph{both} $\<\Psi_E|\hat{C}^{\dag}|\Psi_O\>$ and $\<\Psi_E|\hat{C}|\Psi_O\>$ to be non-zero. In general, this will be satisfied as long as $\<\Psi_E|\chatdag_i|\Psi_O\>$ and $\<\Psi_E|\chat_i|\Psi_O\>$ are both themselves non-zero for $i$ within the region our MBS is localised to.

For our system, we want to consider excitations on either the first or the second site, and thus we calculate $\<\Psi_E|\chatdag_{1\+}|\Psi_O\>=\frac{1}{2\sqrt{2}}\sin(\theta/4)$, $\<\Psi_E|\chat_{1\+}|\Psi_O\>=\frac{1}{2}e^{-i\phi_+/4}\cos(\theta/4)$. Similar results are found for the other spin species and sites. We see that this condition does hold for our system, except at $\theta=0$ (up to rotations of $2\pi$). This is expected, as spin-orbit coupling is known to be required for Majorana bound states to exist \cite{AliceaReview}.

Finally, as in \cite{Kitaev}, we require that our Majorana bound states come in spatially separated pairs (i.e. pairs which are separated by a distance greatly exceeding the exponential localisation length). In our system spatial separation is limited, as we only have two sites. This problem is naturally dealt with in larger arrays of quantum dots, which will be required to demonstrate existence of braiding (which cannot be acheived with only two MBSs \cite{SimonReview}). For experimental realisation of localisation in our system, we need to fine-tune our parameters enough that the localisation length (which is a function of fluctuations in the ground state energy gap) is much less than a single site. \color{black} This is especially important, as unlike the quadratic protection found in the system of \cite{LeijnseFlensberg}, the band gap here grows linearly with our shift from the degeneracy. However, we have great control over the superconducting phase difference $\phi_-$ through the flux $\Phi_S$. For small deviations from the degeneracy ($\Delta E<<E_Z$), we would expect the MBSs to delocalise exponentially over the system in a manner similar to the Kitaev chain \cite{SauDasSarma}. At deviations larger than this, mixing of Kramers pairs will be the biggest concern, as our proposal does not protect this in the way of \cite{WolmsSternFlensberg}.
\color{black}

We can relax the localisation condition somewhat, as suggested in \cite{TonyMennoPaper}. To do so, we note that a braiding consists of evolving a creation operator by a phase $e^{i\phi}$. The corresponding annihilation operator must then evolve by the phase $e^{-i\phi}$, and so the number operator will be invariant. We should thus be able to use any number operators we wish to describe our Majorana bound state, whilst retaining the localization for the purposes of braiding by insisting that single products of creation and annihilation operators are restricted to a small number of neighbouring sites. This argument is true for any spin rotation (as we cannot spatially separate spins). As such, we define the rotated number operators $\hat{n}_{i\sigma\rho}$ by

\begin{align}
\hat{n}_{i\sigma\rho} &= \cos^2(\rho/2)\hat{n}_{i\sigma_z}+\sin^2(\rho/2)\hat{n}_{i\bar{\sigma}_z}\nonumber\\&\hspace{2cm}+\frac{1}{4}\sin(\rho)(\hat{n}_{i\sigma_x}-\hat{n}_{i\bar{\sigma}_x}),
\end{align}

which have corresponding rotated creation operators $\chatdag_{i\sigma\rho}=\cos(\rho/2)\chatdag_{i\sigma}+\sigma\sin(\rho/2)\chatdag_{i\bar{\sigma}}$.

Our Majorana bound state is then a self-adjoint operator that excites each ground state to the other, made up of products of these rotated number operators, and the creation and annihilation operators from a single site. In \cite{TonyMennoPaper} it was required only that we consider the effect of this excitation within the ground state subspace, but it can be generalised to the entire $16$ dimensional Hilbert space. If we set $\rho=\frac{\pi}{2}-\frac{\theta}{2}$ and $\eta=\frac{\pi}{2}+\frac{\theta}{2}$, and define

\begin{align}
\hat{z}_{i\sigma}&=1-\hat{n}_{i\sigma},\\ 
\hat{g}_{i\sigma}&=e^{-i\phi_+/4}\chatdag_{i\sigma}+e^{i\phi_+/4}\chat_{i\sigma},\label{g1Eqn}\\
\hat{g}'_{i\sigma}&=\frac{1}{\sqrt{2}}(e^{i\phi_+/4}\chatdag_{i\sigma}+e^{-i\phi_+/4}\chat_{i\sigma}),\label{g2Eqn}
\end{align}

then we can write operators $\gamma_1$ and $\gamma_2$ localized to site $1$ and $2$ respectively. \color{black} These correspond to quasiparticles with finite lifetimes as discussed in section \ref{sec:EffHam}.
\color{black}

\begin{widetext}
\begin{align}
\gamma_1&=\zhat_{1\-}\{-\hat{g}_{1\+}\cos(\theta/4)\zhat_{2\+\eta}\zhat_{2\-\eta}+\hat{g}'_{1\+}[\sin(\theta/4)(\hat{n}_{1\+\eta}\zhat_{2\-\eta}+\hat{n}_{2\-\eta}\zhat_{2\+\eta})+\cos(\theta/4)(\hat{n}_{2\-\eta}-\hat{n}_{2\+\eta})]\}\nonumber\\
&+\zhat_{1\+}\{\hat{g}_{1\-}\sin(\theta/4)\zhat_{2\+\eta}\zhat_{2\-\eta}+\hat{g}'_{1\-}[\cos(\theta/4)(\hat{n}_{2\+\eta}\zhat_{2\-\eta}+\hat{n}_{2\-\eta}\zhat_{2\+\eta})+\cos(\theta/4)(\hat{n}_{2\-\eta}-\hat{n}_{2\+\eta})]\}\\
\gamma_2 &=\zhat_{2\-}\{\hat{g}_{2\+}\cos(\theta/4)\zhat_{1\+\rho}\zhat_{1\-\rho}+\hat{g}'_{2\+}[\sin(\theta/4)(\hat{n}_{1\+\rho}\zhat_{1\-\rho}+\hat{n}_{1\-\rho}\zhat_{1\+\rho})+\cos(\theta/4)(\hat{n}_{1\+\rho}-\hat{n}_{1\-\rho})]\}\nonumber\\
&+\zhat_{2\+}\{\hat{g}_{2\-}\sin(\theta/4)\zhat_{1\+\rho}\zhat_{1\-\rho}+\hat{g}'_{2\-}[-\cos(\theta/4)(\hat{n}_{1\+\rho}\zhat_{1\-\rho}+\hat{n}_{1\-\rho}\zhat_{1\+\rho})+\sin(\theta/4)(\hat{n}_{1\+\rho}-\hat{n}_{1\-\rho})]\}.
\end{align}
\end{widetext}
 Note that these operators are not unique; for example the operator $\hat{n}_{1\+\rho}\hat{n}_{1\-\rho}$ does not act on $|\Psi_E\>$ or $|\Psi_O\>$ in this limit, and so terms containing this operator can be removed, allowing us to rewrite our Majorana bound state in terms of products of no more then $5$ creation and annihilation operators. These operators specifically were chosen as they act only on the finite energy states, and have other properties which will be discussed in the next section.

\begin{figure}[t]
\begin{center}
\includegraphics[width=8.5cm]{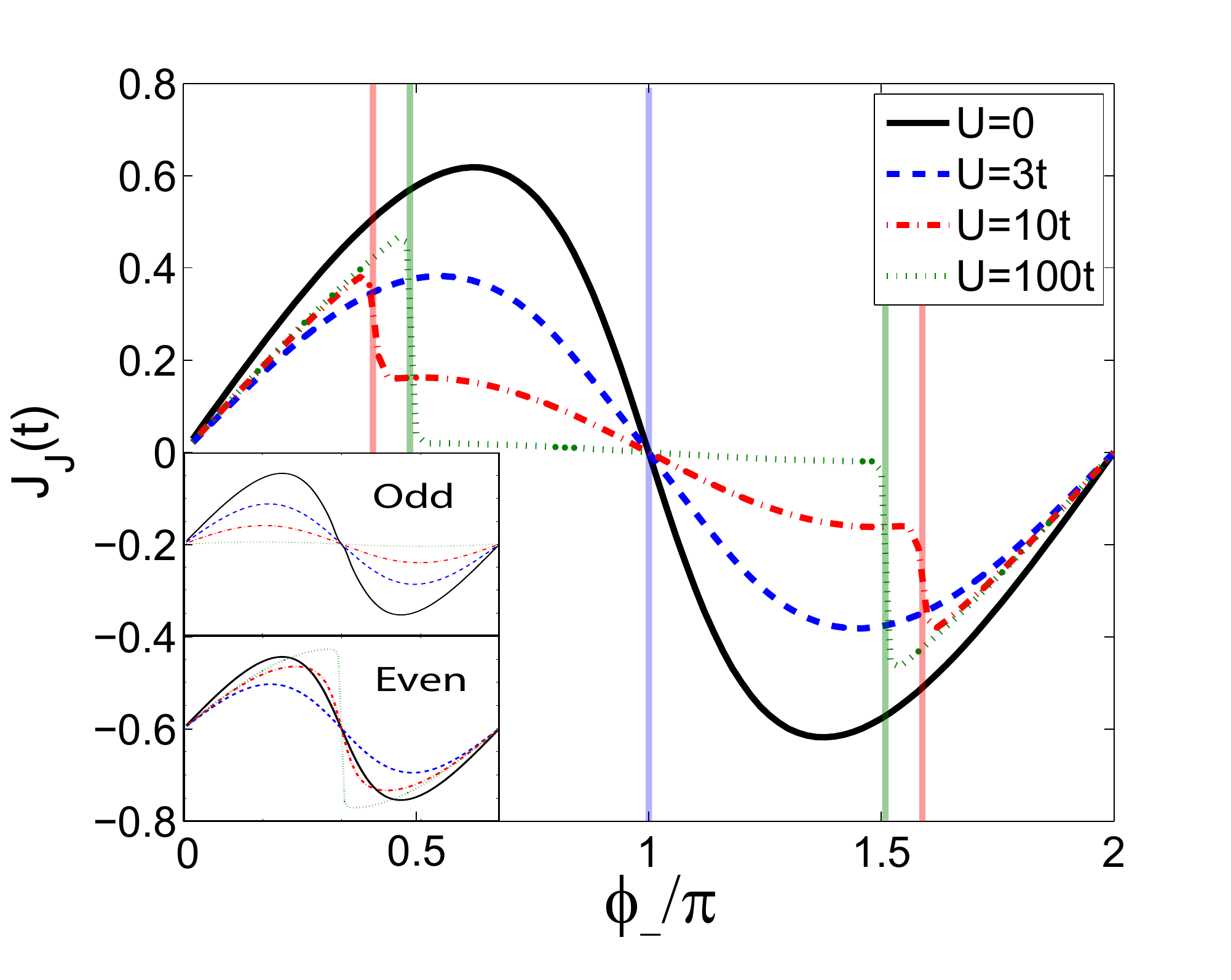}
\caption{(main) Josephson current in the double quantum dot, assuming parity is not conserved. Vertical lines indicate the values of $\phi_-$ required for a ground state degeneracy in the system of matching colour. If a measurement was made slowly enough, fluctuations in the parity of either superconductor would ensure this is the case. (inset) the same plots, but with conservation of parity, assuming we are either in the odd (top) or the even (bottom) sector. A real measurement would likely fall between the two, giving a measure of how well parity is conserved.}
\label{JosephsonNoParity}
\end{center}
\end{figure}

\begin{figure}[t]
\begin{center}
\includegraphics[width=8.5cm]{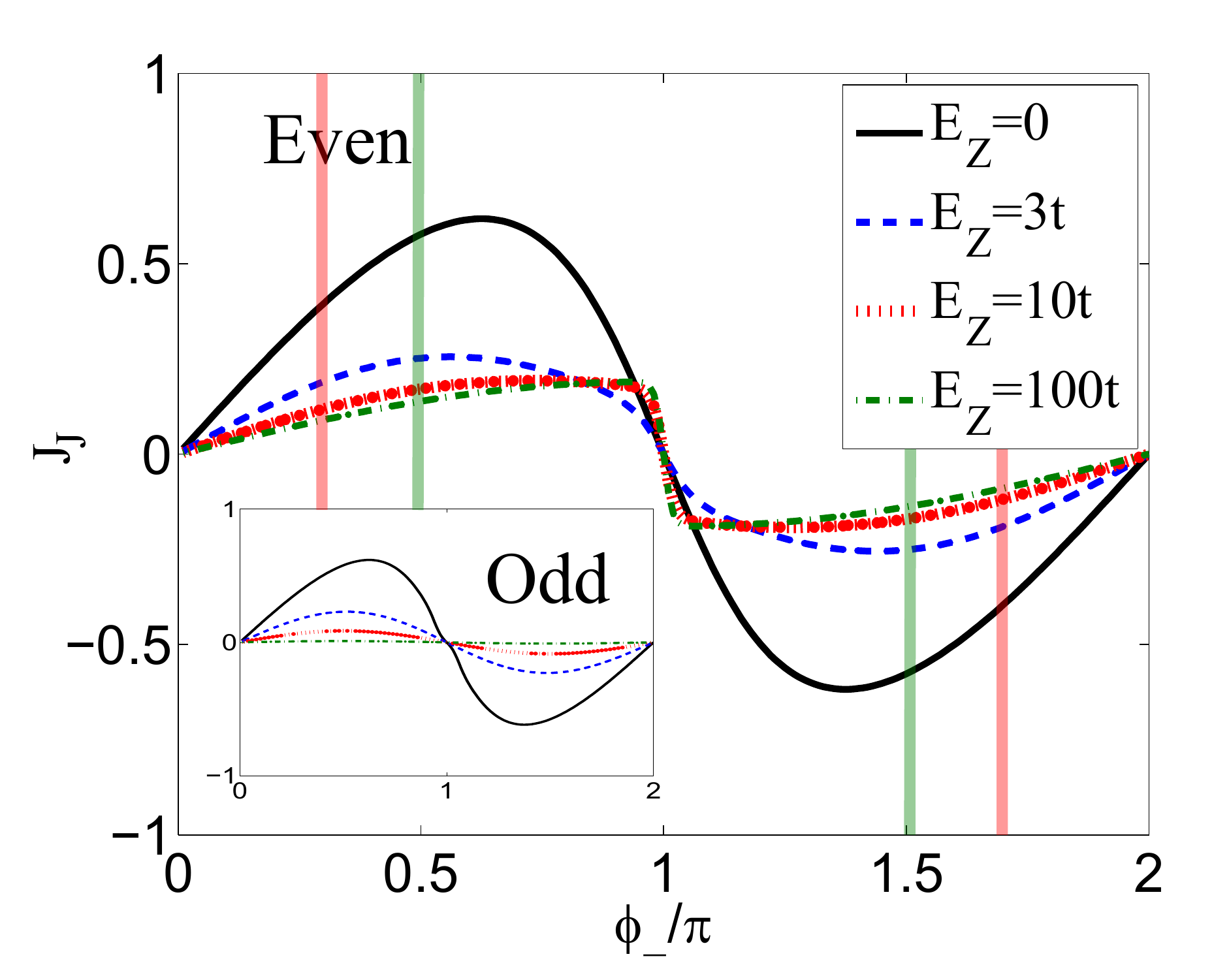}
\caption{Josephson current results for the even particle number sector with increasing $E_Z$, as we head towards the Kitaev regime ($\theta=\pi/2$, $E_Z\rightarrow\infty$). Vertical lines indicate the values of $\phi_-$ required for a ground state degeneracy in the system of matching colour. We see that this has a similar effect to increasing $U$, but the current through the even sector drops to one quarter of the strength. This is due to the Kitaev regime not being able to access as many even particle number states to transmit current. (Inset) the odd particle number sector is similar to the increasing $U$ results also.}
\label{JosephsonZFig}
\end{center}
\end{figure}

In order to provide further evidence that these operators should correspond to Majorana bound states, we demonstrate a method by which our system can be continuously tuned to the one-dimensional wire model of Kitaev. If our on-site energies are locked to the energy of the Zeeman field ($\epsilon_1+t=\epsilon_2+t=E_Z$), then we effectively have a zero-energy on-site potential for spin-up electrons, and an on-site potential for spin down electrons equal to $2E_Z$. In the limit as $E_Z\rightarrow\infty$, this removes the possibility of spin down excitations. An effective Hamiltonian can then be written for the remaining states by removing all terms that contain creation or annihilation operators for spin down electrons, leaving

\begin{figure*} [t]
\begin{center}
\begin{tabular}{cc}
\includegraphics[width=8.5cm]{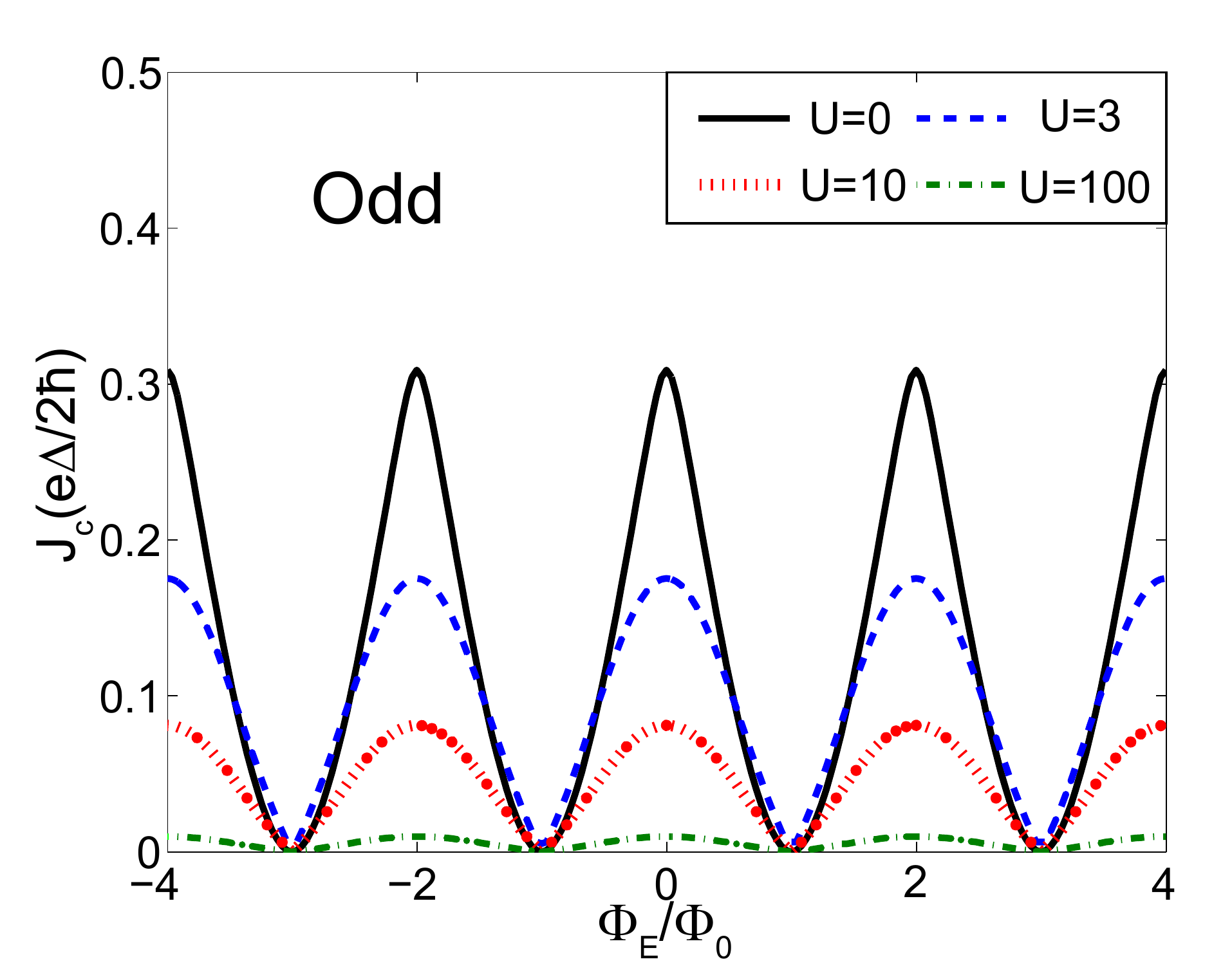}&
\includegraphics[width=8.5cm]{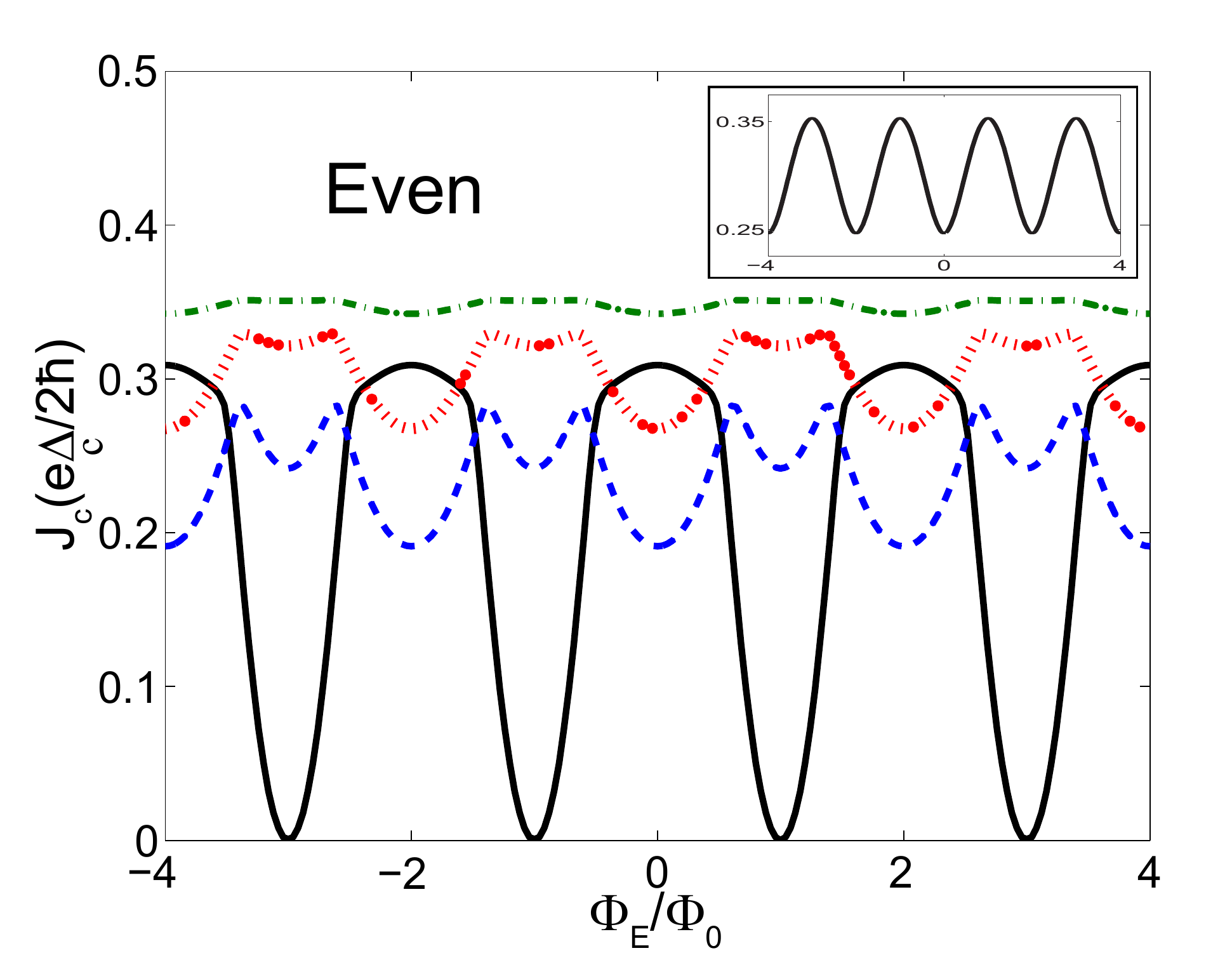}
\end{tabular}
\caption{The double slit-like interference pattern made by plotting the critical Josephson current as a function of the magnetic flux through the quantum dot Josephson junction for the odd and even number parity sectors (labeled). We see that this saturates as the electron-electron interaction term grows stronger, which can be explained by the current being carried by chargeless particles. (Inset) the critical current assuming that the parity is not conserved. This pattern is caused by the changing of the relative energies of the ground even and odd particle number parity eigenstates, and is not due to particle interference.}
\label{DoubleSlitDiffraction}
\end{center}
\end{figure*}

\begin{align}
\Hh_{eff}&=t\cos(\theta/2)\cos(2\Lambda)\chatdag_{1\+}\chat_{2\+}\nonumber\\&-\Delta e^{i\phi_+/2}\cos(\phi_-/2)\sin(\theta/2)\chatdag_{1\+}\chatdag_{2\+}+\text{h.c.}
\end{align}

We see that this is the superconducting wire model of Kitaev for two sites. This system will then support Majorana bound states when $t\cos(\theta/2)\cos(2\Lambda)=\pm\Delta \cos(\phi_-/2)\sin(\theta/2)$. In Fig. \ref{fig:ChangingPhase}, we demonstrate the possibility to tune between this limit and the previously considered system with vanishing $E_Z$, whilst retaining a degeneracy at all times. This provides further evidence for the existence of Majorana bound states in the small $E_Z$ limit, \color{black} as we do not see any evidence for a phase transition during this tuning.\color{black}

If the large $E_Z$ limit could be achieved in an experiment, it may have some advantages over wire systems in which the signatures of Majorana bound states were previously measured. As the magnitude of the effective order parameter can be tuned, it should be relatively easy to find a region where Majorana bound states are present. Also, as our dots are discrete, we should hopefully be able to measure the localization of Majorana bound states in a line of dots (where a similar limit presents itself) to the ends, as others should have minimal conductance.

\section{\label{sec:Josephson} Current-phase relationship and Fraunhofer pattern}
In this section we discuss the Josephson supercurrent through the double quantum dot. Specifically, we study the dependence of the supercurrent on the phase difference $\phi_-$, and its behaviour when the magnetic flux passing between the dots is changed. The Josephson supercurrent can be calculated as the derivative of the free energy with respect to the superconducting phase difference \cite{Droste}$J=\partial_{\phi_-}-T\ln\sum_i e^{-E_i/T}$. In the limit of infinite $U$, supercurrent is absent in the odd parity state, as both local and crossed Andreev reflection are not possible \cite{TonyMennoPaper}. By comparison, a Josephson current exists for the even number parity sector, and is $4\pi$ periodic in the absence of relaxation. A measurement of the periodicity of the Josephson current could then be used to determine how well parity is conserved. If we start at some value of $\phi_-$ at which the system is non-degenerate, and tune it through a degeneracy, we shift the energy levels until they cross. If our system was perfectly conserving of parity, our now-excited state would not be able to relax into the new ground state, and the Josephson current would either be $4\pi$ periodic or flat, depending on the initial state of the system. However, as we are still connected to the superconducting leads, we would realistically expect some perturbation of these to occur after a finite period of time that would break this parity conservation. This would then correspond to a sudden jump to the ground state, and a corresponding change in the Josephson current. As the perturbation frequency increases, the free energy would become a function of the entire system rather than one parity sector. In Fig. \ref{JosephsonNoParity}, we plot this for various values of $U$ (with parity-conserving counterparts inset, and vertical lines to indicate the values of $\phi_-$ required for a ground state degeneracy). Measurements of deviations from this plot as the sweep time (across $\phi_-$) is decreased would then give a measure of the coherence time of our system's parity conservation.

If we measure the Josephson current as we increase $E_Z$ whilst holding $\epsilon_i=E_Z$ (which takes us towards the effective Kitaev wire described previously), we find that the behaviour mimics that for increasing $U$, except for one difference. At large $E_Z$, all four doubly occupied states are gapped out save for one (where both electrons are spin-up), and our current is thus reduced four-fold. This is seen in Fig. \ref{JosephsonZFig}. The similarity in behaviour between increasing $U$ and increasing $E_Z$ provides further evidence that the Majorana bound state picture is correct for the large $U$ limit, as it has similar characteristics to the large $E_Z$ limit where Majorana bound states are expected to appear.

In Fig. \ref{DoubleSlitDiffraction} we show the dependence of the supercurrent on the magnetic flux piercing the Josephson junction (see Fig. \ref{fig:Schematic}). The standard Fraunhofer pattern \cite{FraunhoferPaper} arises from interference of the supercurrent density in a junction, which can be thought of as a single slit of finite width. By contrast, we are considering a magnetic flux tube between the dots, and only allowing electrons to pass through either dot. This can be roughly considered a double-slit for electron transport. As such, we expect a double slit-like pattern in the critical current. We show this in Fig. \ref{DoubleSlitDiffraction} for both the even and odd sectors, and the entire system. 

As $U$ is increased, we see that the diffraction patterns for either number parity sector saturate, but at different levels. This is to be expected, as the even ground state permits crossed Andreev reflection only, which has no dependence on the flux between the dots, whereas the odd ground state permits elastic co-tunneling only, which does not permit Josephson current. However, it can also be explained by tunneling through the Majorana bound states. To see this, consider the product $\gamma_1\gamma_2$ of our Majorana operators. This can be written as the sum of two parts, $(\gamma_1\gamma_2)_O+(\gamma_1\gamma_2)_E$, which act on the odd and even sectors of the Hilbert space individually. Each term in the odd sector must contain a lone creation or annihilation operator from each site with either spin, and as our operators only act on the finite energy states, where fermion number is conserved, all terms will take the form of $\chatdag_1\chat_2$ (or the Hermitian conjugate), multiplied by an appropriate number operator. This excitation then describes only elastic co-tunneling on the odd states. For the even states, we calculate explicitly
\begin{align}
(\gamma_1\gamma_2)_E&=\frac{e^{i\phi_+/2}}{\sqrt{2}}(\cos(\theta/2)(\zhat_{1\-}\zhat_{2\+}\chat_{2\-}\chat_{1\+}+\zhat_{1\+}\zhat_{2\-}\chat_{1\-}\chat_{2\+})\nonumber\\&+\sin(\theta/2)(\zhat_{1\-}\zhat_{2\-}\chat_{2\+}\chat_{1\+}+\zhat_{1\+}\zhat_{2\+}\chat_{2\-}\chat_{1\-}))+\text{h.c.}
\end{align}
We see here that the excitation describes only crossed Andreev reflection for the even states. The Majorana picture then gives the expected result for Josephson current, justifying it further.

The Josephson current through the total system (without parity conservation) displays an interesting trait here, as it retains the double-slit pattern, but picks up a $\pi$ phase shift at $U\rightarrow\infty$. This is due to the Josephson current in the even number parity ground state being the highest when the state is the highest energy, which then requires the odd number parity ground state to be higher energy still. This energy is $\Phi_J$-dependent in the manner shown above. As such, this is not a pattern caused by interference between electrons.

\section{\label{sec:conclusion}Conclusion}
We have investigated the double quantum dot model, first proposed in \cite{TonyMennoPaper} as a potential system to support Majorana bound states. We have derived an effective Hamiltonian of the system, and modeled the spectra, showing a range of parameters for which the system should be easily tunable to a degeneracy. In the limit as $U\rightarrow\infty$, we have written down the form of a Majorana bound state that excites between the ground states. We have shown that the degeneracy can be tuned continuously to a system equivalent to the one-dimensional wire model of Kitaev. Finally, we have discussed how measurements of the Josephson current can display the conservation of parity in the system, and a measurement of the Fraunhofer-type effect associated with Josephson current that disappears when current travels through chargeless modes.

To the best of the authors' knowledge this model presents the first example of Majorana bound states in a system with strong correlations, \color{black} where they do not present themselves as single-particle excitations\color{black}. This makes it important to justify the fact that these excitations are Majorana-like in nature. The similarities to the Kitaev model, and the existence of a continuous transition to this model presents a strong case, which is backed up by the results from the Josephson current. While the two-dot setup as proposed here is the simplest system to construct Majorana bound states, observing non-Abelian statistics from the many-particle Majorana bound states demands larger systems to move the Majorana bound states around each other. These larger systems will also provide topological protection as the Majorana modes are separated by the number of dots \cite{SauDasSarma, Fulga2013}. 
\color{black}

\acknowledgements{We would like to thank Ben J. Powell for enlightening discussions. ARW is financially supported by a University of Queensland Postdoctoral Research Fellowship. MV is financially supported by the Australian Research Council Centre of Excellence for Quantum Computation and Communication Technology (Project No. CE11E0096), the U.S. Army Research Office (Grant No. W911NF-13-1-0024), and the Netherlands Organization for Scientific Research (NWO) by a Rubicon Grant.}

%

\end{document}